\documentclass[twocolumn, showpacs, showkeys, pra, superscriptaddress]{revtex4-1}

\usepackage{amssymb,amsmath,xcolor,graphicx,xspace,colortbl,ragged2e,rotating} %
\usepackage{units} 
\DeclareGraphicsExtensions{.pdf,.eps,.ps,.png,.jpg,.jpeg}

\begin{document}
\title{Infinite-randomness fixed point of the quantum superconductor-metal transitions in amorphous thin films
}
\author{Nicholas A. Lewellyn}
\affiliation{
School of Physics and Astronomy, University of Minnesota, Minneapolis, MN 55455, USA}
\author{
Ilana M. Percher}
\affiliation{
School of Physics and Astronomy, University of Minnesota, Minneapolis, MN 55455, USA}
\author{
JJ Nelson}
\affiliation{
School of Physics and Astronomy, University of Minnesota, Minneapolis, MN 55455, USA}
\affiliation{
Current address: Department of Physics, University of Rochester, Rochester, NY 14627, USA}
\author{Javier Garcia-Barriocanal}
\affiliation{
Characterization Facility, University of Minnesota, Minneapolis, MN 55455, USA}
\author{
Irina Volotsenko}
\affiliation{
Department of Physics, Bar Ilan University, Ramat Gan, ISRAEL}
\author{
Aviad Frydman}
\affiliation{
Department of Physics, Bar Ilan University, Ramat Gan, ISRAEL}
\author{Thomas Vojta}
\affiliation{Department of Physics, Missouri University of Science and Technology, Rolla, Missouri 65409, USA}
\author{Allen M. Goldman}
\affiliation{
School of Physics and Astronomy, University of Minnesota, Minneapolis, MN 55455, USA}

\date{
\today
}
\begin{abstract}
The magnetic-field-tuned quantum superconductor-insulator transitions of disordered amorphous indium oxide films are a paradigm in the study of quantum phase transitions, and exhibit power-law scaling behavior.  For superconducting indium oxide films with low disorder, such as the ones reported on here, the high-field state appears to be a quantum-corrected metal.  Resistance data across the superconductor-metal transition in these films are shown here to obey an activated scaling form appropriate to a quantum phase transition controlled by an infinite randomness fixed point in the universality class of the random transverse-field Ising model. Collapse of  the field-dependent resistance vs. temperature data is obtained using an activated scaling form appropriate to this universality class, using values determined through a modified form of power-law scaling analysis. This exotic behavior of films exhibiting a superconductor-metal transition is caused by the dissipative dynamics of superconducting rare regions immersed in a metallic matrix, as predicted by a recent renormalization group theory.  The smeared crossing points of isotherms observed are due to corrections to scaling which are expected near an infinite randomness critical point, where the inverse disorder strength acts as an irrelevant scaling variable.
\end{abstract}
\maketitle

\section{Introduction}

The magnetic-field-tuned quantum superconductor-insulator transition (SIT) of quasi-two-dimensional amorphous indium oxide thin films have been studied for almost three decades (for early examples see Refs.\ \cite{HebPal, Yaz}). Generally, superconductor-insulator transitions can be tuned in several ways such as with perpendicular and parallel magnetic fields, charge carrier concentration, or disorder~\cite{Lin}. The nature of these quantum phase transitions is not settled. The canonical theory for the perpendicular field-tuned superconductor-insulator transition implies that the transition is directly from insulator to superconductor, without an intermediate metallic regime. A finite, nonzero resistance is expected only at the quantum critical point (QCP), which is predicted to have a universal resistance value of $h/4e^{2} $~\cite{FisherM}. Experimental observations of broad metallic regimes between the superconducting and insulating regimes have been reported, seemingly contradicting this prediction~\cite{Qin, Mason, failed}. In two-dimensional crystalline films  quantum superconductor-to-metal transitions (SMTs)~\cite{Tsen} have also been reported, and have been interpreted as evidence of a Bose metal~\cite{Das1, Das2, Dalid1, Dalid2}. However in some instances it is difficult to prove that these metallic regimes are not just artifacts caused by heating due to the measuring current, radio frequency interference, or some other source, as disordered superconducting thin films are extremely sensitive to external perturbations~\cite{Tamir2018}.

Quantum phase transitions occur at zero temperature when the ground state of a system changes in response to a variation of parameters in the Hamiltonian. Since zero absolute temperature is experimentally inaccessible, the presence of such a transition must be inferred from changes of measurable properties that are influenced by quantum fluctuations that persist to nonzero temperatures. In the case of superconductor-insulator transitions, film resistance measurements are commonly analyzed using scaling.  The resistance of disordered superconducting films near a magnetic-field-tuned superconductor-insulator transition can be described in terms of a power-law scaling form~\cite{SGCS}
\begin{equation}
R \left(\delta ,T \right) =\Phi \left (\delta T^{ -1/\nu z}\right)
\label{eq:FSS}
\end{equation}
where $\delta  = \left \vert B -B_{c}\right \vert /B_{c}$ is the distance from the critical field $B_{c}$ and $\Phi$ is a scaling function.
This scaling form implies that the magnetoresistance isotherms ($R$ vs. $B$ curves at fixed $T$) all cross at the critical field $B_c$.
Moreover, the magnetoresistance isotherms are expected to collapse into two branches when plotted as function of $\delta T^{ -1/\nu z}$ for the correct value of the exponent product $\nu z$. Here, $\nu $ is the correlation length exponent and $z$ is the dynamical critical exponent. In principle, knowledge of these exponents can be used to identify the universality class of the transition.

The electrical transport data for the lower-resistance indium oxide films studied here do not fall neatly into this description. The
high-magnetic-field regime is metallic rather than insulating, which is a consequence of these lower-resistance films being less disordered than films that exhibit a direct superconductor-insulator transition. Instead of a single magnetic field at which magnetoresistance isotherms cross, a series, or essentially a continuum, of crossing fields is observed. Similar effects were reported by Gantmakher and collaborators two decades ago~\cite{Gantm2000} and at the time were analyzed using an ad hoc scaling form.

In a number of recent publications, the systematic variation of the crossing field with temperature was found to be accompanied
by a strong systematic variation of the value of the effective exponent $\nu z$ determined at each crossing point. Examples include the superconductor-metal quantum phase transitions of ultrathin single crystal Ga films~\cite{xing}, La\textsubscript{2}AlO\textsubscript{3}/SrTiO\textsubscript{3} interfaces~\cite{Shen}, ionic liquid-gated single-crystal flakes of ZrNCl and MoS\textsubscript{2}~\cite{Saito2018}, and of monolayer NbSe\textsubscript {2}~\cite{XingN}.
The analysis employed in these works involved the use of the power-law scaling form (\ref{eq:FSS}) at crossing points at selected temperatures, using nearby isotherms to collapse the data and determine effective values of the exponent product $\nu z$ as a function of temperature. These effective values were found to diverge as the quantum phase transition is approached, i.e., for $T\to 0$ and $B\to B_c$.
This behavior was interpreted as being evidence of a quantum Griffiths singularity \cite{Grif,ThillHuse95,YoungRieger96} associated with an infinite-randomness critical point \cite{Fisher95, Motrunich}, as had been predicted by a renormalization group calculation \cite{Hoyos2007,Vojta2009} for a quantum superconductor-metal phase transition (for reviews, see, e.g., Refs.\ \cite{Vojta06,VojtaRev}).

This theory also predicts that a quantum superconductor-metal phase transition governed by an infinite randomness fixed point features activated dynamical scaling, rather than power-law scaling. In this case, $z \rightarrow \infty$ as $T \rightarrow 0 $, so the scaling form of the resistance differs from Eq.~\ref{eq:FSS}, taking the form \cite{DRHV10}
\begin{equation}
R\left (\delta  , \ln{ \frac{T_{0}}{T}} \right  ) =\Phi \left [\delta \left (\ln \frac{T_{0}}{T}\right )^{1/\nu \psi }\right ],
\label{eq:activated}
\end{equation}
where once again $\delta  =\vert B -B_{c}  \vert /B_{c}$ is the distance from the critical field, and $\nu$ is the correlation length exponent. The exponent $\psi $ is the tunneling exponent, and $T_{0}$ is a microscopic temperature scale, which acts as an additional fitting parameter.  Equation~\ref{eq:activated} predicts a single crossing point in magnetic field, and does not account for the temperature-dependence of the crossing fields observed here.  The smeared crossing points result from corrections to scaling which become less important as the temperature is decreased toward zero. These are explained in the appendix.

We will show in the present work on indium oxide films exhibiting superconducting-metal transitions that curves of resistance vs. temperature, $R(T)$, at different magnetic fields, of films with smeared crossing points of magnetoresistance isotherms can be collapsed using activated scaling (Eq.~\ref{eq:activated}). This provides strong evidence for a quantum superconductor-metal phase transition governed by an infinite randomness fixed point with activated dynamical scaling. Our paper is organized as follows. In Sec.~\ref{Sec:Exp}, we briefly describe the experimental methods. Section \ref{Sec:Res} presents the experimental results. We describe the scaling analysis in Sec.~\ref{Sec:Ana}, paying particular attention
to the relationship between power-law and activated dynamical scaling. We conclude in  Sec.~\ref{Sec:Con}
by putting our results into a broader perspective.

\section{Experimental Methods}
\label{Sec:Exp}

The InO\textsubscript {x } films used for these studies were about \unit[30]{nm} thick, and were grown by electron beam evaporation of  In\textsubscript {2}O\textsubscript {3}. During deposition, an O\textsubscript {2} partial pressure between $2 \times 10^{-5}$  and \unit[$9 \times 10^{ -4}$]{mbar} was maintained in the chamber by bleeding gas through a needle valve while continuing to pump~\cite{OVAD}. Amorphous films were produced when the substrate temperature was kept below about $40^{\circ}$C.  These films then sat at ambient temperature in air for about three years, during which time they underwent annealing. This process does not change the carrier concentration, but reduces the disorder.  Subsequent measurements were initially carried out using a Quantum Design Physical Properties Measurement System to determine the basic characteristics of the films and then with an Oxford Kelvinox-25 dilution refrigerator for lower temperature and detailed measurements.

The range of temperatures over which these measurements are reliable is limited by factors such as electromagnetic noise, self-heating due to the measuring current, and limitations of the cooling power and base temperature of the dilution refrigerator employed.  The leads to the cryostat were filtered only at room temperature, so that there was electromagnetic noise delivered to the sample.  Measurements of resistance were confined to currents at which the I-V characteristics were linear, eliminating the possibility of heating due to the measuring current.

The minimum achievable temperature at which the data was reliable was determined from the behavior of the high-field metallic regime above the transition.  The conductance in this regime, if it corresponds to that of a 2D quantum corrected metal should be a linear function of the natural logarithm of temperature~\cite{Alt, Kramer}.  The temperature at which the conductance deviated from this form at high magnetic fields was then taken as the minimum temperature at which reliable measurements and analysis could be carried out.

\section{Results}
\label{Sec:Res}

\begin{figure}
\centering
\includegraphics[width=0.46\textwidth ]{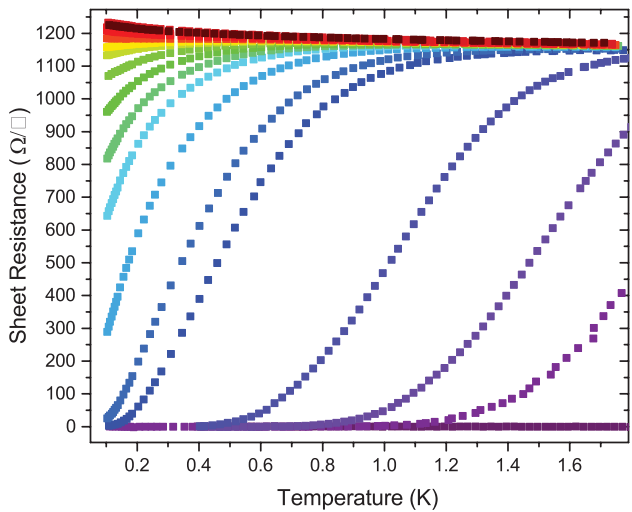}
\caption{(Color online) Sheet resistance vs. temperature at magnetic fields of 0, 3.0, 4.0, 5.0, 6.0, 6.2, 6.5, 6.7, 6.8, 6.9, 7.0, 7.100, 7.150, 7.225, 7.325, 7.4, 8.0, and \unit[12.0]{T} (bottom to top). }
\label{fig:RvsT}
\end{figure}

The InO\textsubscript {x} films studied
exhibited  zero-field transition temperatures of approximately \unit[2.8]{K}.  Curves of resistance $R$ vs.\ temperature $T$ of one of the films at various magnetic fields $B$ are shown in Fig.~\ref{fig:RvsT}.
At perpendicular magnetic fields $B \approx \unit[7]{T}$, the temperature dependence $dR/dT$ of the resistance changes sign. This change occurs at a resistance that is much lower than the quantum resistance $h/4e^{2} $ for Cooper pairs (which is the typical value for a direct superconductor-insulator transition).

\begin{figure}
\centering
\includegraphics[width=0.46\textwidth ]{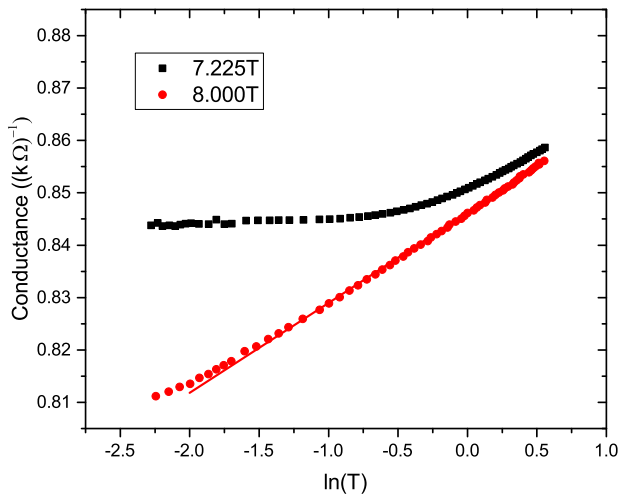}
\caption{(Color online) Conductance vs. $\ln {T}$ in fields of 7.225 and \unit[8]{T}.  The coefficient of  $\ln{T}$ for the linear fit in units of $e^{2}/h$ is 0.4435.  Conductance varying linearly with $\ln {T}$ at fields above those at which a crossover is found (see Fig.~\ref{fig:RvsH}) is a clear indication of a quantum corrected metal.}
\label{fig:Metalfits}
\end{figure}

\begin{figure}
\centering
\includegraphics[width=0.46\textwidth ]{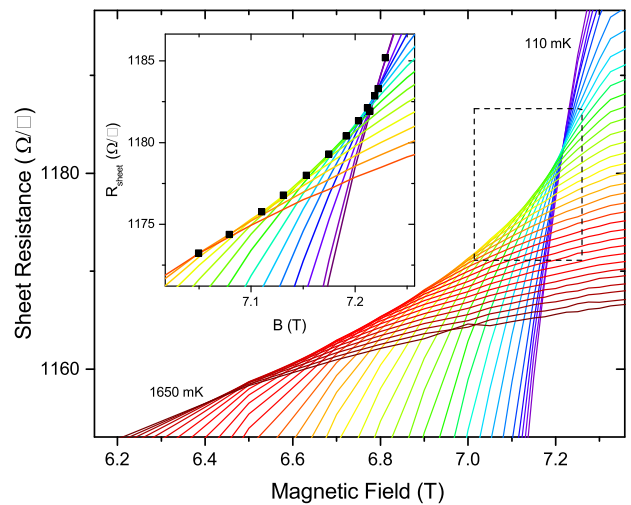}
\caption{(Color online) Detailed view of the magnetoresistance isotherms close to the quantum phase transition, showing a series or continuum of crossings spread out over a range of temperatures and fields. The temperatures shown are \unit[$110-170$]{mK} in \unit[20]{mK} steps and \unit[$200-1700$]{mK} in \unit[50]{mK} steps
(top to bottom on the r.h.s. of the plot). The inset shows isotherms at temperatures of 160 mK, 180 mK and 200-800mK in 50 mK steps. The crossing points are marked by black squares. the region shown in the inset is marked on the main plot with a dashed box.}
\label{fig:RvsH}
\end{figure}

The films exhibited metallic behavior under magnetic fields greater than \unit[8]{T}, as signified by the
linear dependence of their conductances on the logarithm of temperature (see Fig.~\ref{fig:Metalfits}).
This is the expected behavior for a conventional 2D quantum-corrected disordered metal \cite{Alt, Kramer}.
Additionally, there was what might be termed a metallic regime at magnetic fields intermediate between those in which the films were obviously superconducting and those in which they were metallic. In this regime the values of $dR/dT$ were positive, suggesting the onset of superconductivity, however their resistances did not fall to zero at the lowest measurable temperatures. In subsequent analysis we will assume that a film in this field range with perhaps the exception of the highest fields in the range, would ultimately become superconducting.
Magnetoresistance isotherms were generated using the measured $R(T,B)$ curves by carrying out a matrix inversion of the temperature swept data.  At first glance, it appeared that there was a single crossing point as would be typical for a conventional quantum superconductor-insulator transition. However, a detail of the crossing region, displayed in Fig.~\ref{fig:RvsH} reveals that there is a series (or continuum) of crossings, spread out over a range of temperatures and magnetic fields.
The crossing magnetic fields increase with decreasing temperature and appear to saturate in the limit of zero temperature. This unusual phenomenology is not compatible with the standard power-law scaling analysis. As will be discussed in the next section, it can be explained by activated scaling when subleading corrections to scaling are included.

\section{Scaling analysis}
\label{Sec:Ana}

\begin{figure}
\centering
\includegraphics[width=0.4\textwidth ]{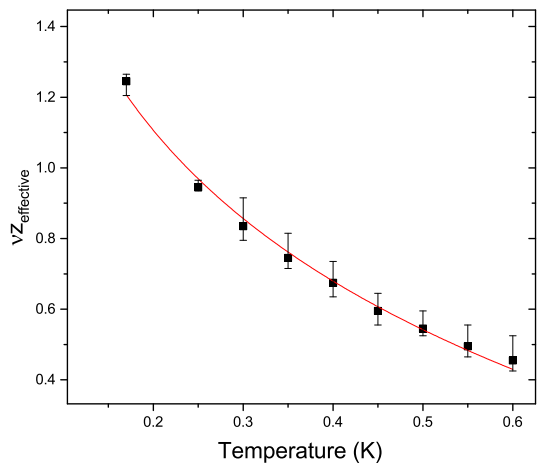}
\caption{(Color online) Effective exponent $(\nu z)_{\rm eff}$ vs. temperature. The solid line is a two-parameter fit to the data of Eq.~\ref{eq:NuZ_eff} with $(\nu \psi)_{\rm eff}$ and $T_{0}$ as adjustable parameters yielding  $( \nu \psi) _{\rm eff}  =0.62$ and $T_{0} =$ \unit[1.21]{K}.}
\label{fig:NuZ}
\end{figure}

We first follow the approach of~\cite{xing}, in which power law scaling is applied to each crossing point to obtain an effective value of the exponent product, $\nu z$, which will be temperature dependent. At a conventional quantum phase transition, the effective values of $\nu z$ are expected to be
constant or at least to saturate at a finite asymptotic value at the critical point.  In contrast, the  $\nu z$ values in Fig.\ \ref{fig:NuZ} increase rapidly as the quantum phase transition is approached, suggesting unconventional behavior.

An important issue in the quantitative analysis is the relationship between the activated dynamical scaling (Eq.~\ref{eq:activated}) expected at an infinite-randomness critical point and the power-law scaling employed in the standard techniques. In the appendix of this paper we show that if a system is governed by activated scaling with corrections to scaling, then the effective value of the exponent product $\nu z$ obtained from a power-law scaling analysis in the vicinity of crossing points found at different temperatures is given by
\begin{equation}
\left( \frac{1}{\nu z} \right) _{\rm eff} =\left( \frac{1}{\nu \psi } \right) _{\rm eff} \frac{1}{\ln ({T_{0}}/{T})}.
\label{eq:NuZ_eff}
\end{equation}
where $( \nu \psi) _{\rm eff}$ is the exponent product for the universality class of the quantum phase transition exhibiting activated scaling. Here again, $\nu$ is the correlation length exponent of the transition and $\psi$ is the tunneling exponent.

The relationship between $1/( \nu \psi) _{\rm eff}$ and the asymptotic value $1/ \nu \psi$ is given by
\begin{equation}
\left( \frac{1}{\nu \psi } \right) _{\rm eff} = \frac{1}{\nu \psi} - \frac{a \omega}{\psi} \left( \ln{ \frac{T_0 }{ T } } \right) ^{- \nicefrac{ \omega}{\psi} }
\label{eq:eff_nu_psi}
\end{equation}
where the corrections in the second term vanish as $T \rightarrow 0$. Here $\omega$ is the leading irrelevant exponent (whose value is not fixed by the existing theories), and the prefactor $a$ is defined in the appendix.

To find values of the effective exponent product $\nu z$ at a crossing point, we considered a sequence of narrow temperature intervals such that the magnetoresistance isotherms within each of the intervals have a well-defined crossing field $B_x(T)$. For the sets of isotherms within each interval a standard power law scaling analysis was performed, collapsing them into one another around their crossing fields $B_x(T)$.  In this case it was important to quantify the extent to which the curves collapsed. This was done by limiting the analysis to points near the crossing, where the scaling function $\Phi$ can be approximated as linear.  The isotherms are plotted as $\ln {R}$ vs.\ $\delta T^{-1/\nu z}$ for a set of possible $\nu z$ values. The upper and lower branches of the rescaled curves are fit to lines, and the $\nu z$ value is chosen for which the upper and lower branches of the curves are both closest to co-linear.  This value that best collapsed the isotherms in a given temperature interval was assigned a temperature equal to the average temperature of the isotherms in this interval.  With this technique, effective values of $\nu z$ as a function of temperature could be found. These values are presented in Fig.\ \ref{fig:NuZ}.

The expression on the right hand side of Eq.~\ref{eq:NuZ_eff} vanishes in the limit of zero temperature, implying that the effective  $\nu z$  diverges. By means of Eq.~\ref{eq:NuZ_eff}, the temperature dependence of the exponent product $\nu z$ obtained from power-law scaling can be used to determine the product $( \nu \psi) _{\rm eff}$ of the activated scaling form (Eq.~\ref{eq:activated}).  The solid line in Fig.~\ref{fig:NuZ} is the result of a two-parameter fit of the data to Eq.~\ref{eq:NuZ_eff}.

The best fit yields an exponent product $( \nu \psi) _{\rm eff}  =0.62$, in good agreement with the numerical predictions \cite{Vojta2009a, Kovac2010, Maest2008} for a two-dimensional infinite-randomness critical point in the random transverse-field Ising universality class. The range of temperatures covered in the analysis of $\nu z$ shown in Fig.~\ref{fig:NuZ} does not extend to a low enough values to make an absolute claim of $\nu z$ divergence in the limit of zero temperature, but fits to a curve that diverges in this limit. This fit further supports the quantum critical point being an infinite-randomness fixed point.

\begin{figure}
\centering
\includegraphics[width=0.48\textwidth ]{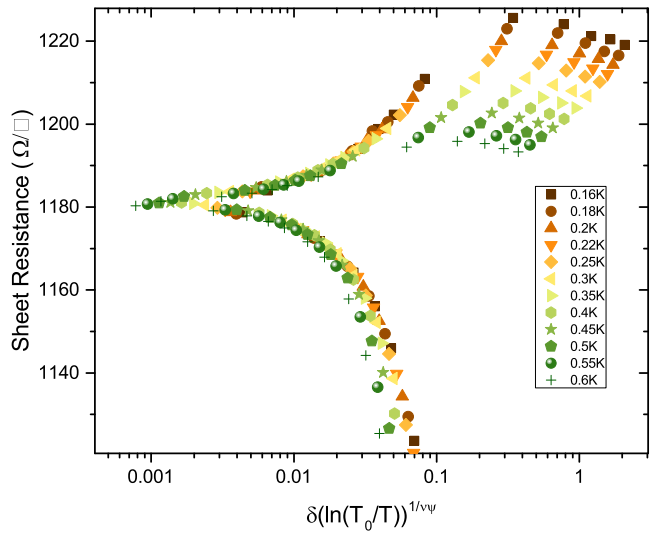}
\caption{(Color online) Sheet Resistance versus the scaling parameter described in Eq.~ \ref{eq:activated} for activated scaling.  The data collapse around critical field $B_{c} =$ \unit[7.21]{T}, with $\nu \psi  =0.62$ and  $T_{0} =$ \unit[1.21]{K} as determined from the fit in Fig.~\ref{fig:NuZ}. }
\label{fig:activated}
\end{figure}

One can solidify this conclusion by scaling the full set of resistance isotherms using Eq.~\ref{eq:activated}, the activated scaling form. At this point in the analysis the only unknown parameter is the critical field of the quantum phase transition, $B_{c}$. To find $B_{c}$ we employed a numerical method used by~\cite{Skinner2018}, in which the variance of the magnetoresistance isotherms plotted against the scaling parameter was minimized.  It was found that a value of $B_{c} =$ \unit[7.21]{T} resulted in the best collapse. This is shown in Fig.~\ref{fig:activated}. Similar results were found when magnetoresistance isotherms from another sample were scaled using the same form (Eq.~\ref{eq:activated}).

While this method gave a well-defined best value for $B_c$, best values for $\nu \psi$ and $T_{0}$ were not easy be determine.  For a fixed value of $B_{c}$, the variance as a function of $\nu \psi$ and $T_{0}$ did not have a well-defined minimum.  Instead there was an extended region in which the variance was roughly minimized.  The values of $\nu \psi$ and $T_{0}$ from the fit to Eq.~\ref{eq:NuZ_eff} fall within this region and yield a reasonable scaling collapse.  It is not surprising that a unique value of $\nu \psi$ could to be determined since it was assumed to be constant.  Within this method, Eq.~\ref{eq:activated} was used to scale the magnetoresistance isotherms and the expected weak temperature dependence of $( \nu \psi) _{\rm eff}$ was not taken into account.

Examining Fig.~\ref{fig:activated}, we see that the scaling collapse breaks down at large values of the scaling parameter for both the upper and lower branches.  In the upper branch, the five disconnected regions correspond to magnetic fields of $8$, $9$, $10$, $11$, and \unit[12]{T}.  Similarly there is a breakdown in the lower branch at \unit[6.7]{T}.  We believe that these breakdowns occur because at low fields and sufficiently low temperatures the film is in an ordered superconducting state not influenced by quantum fluctuations of the order parameter. Correspondingly, at high fields, it is in a metallic state similarly not influenced by quantum fluctuations. The breakdown of scaling in the upper branch occurs at the magnetic field at which the conductance becomes a linear function of the logarithm of the temperature. Thus the breakdown of scaling mark the leaving of the regime of quantum critical behavior, where the scaling is expected to apply.

\begin{figure}
\centering
\includegraphics[width=0.4\textwidth ]{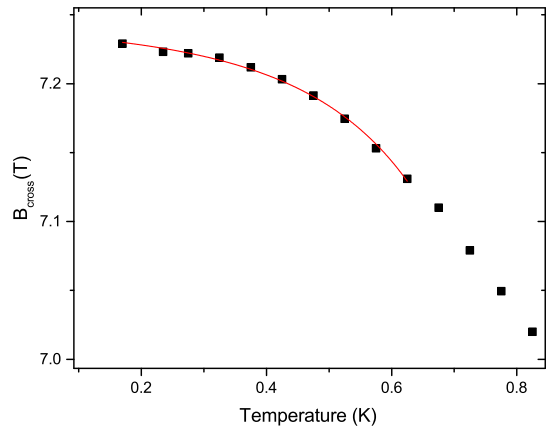}
\caption{(Color online) Crossing field versus temperature.  The solid line is a fit to Eq.~\ref{eq:delta_crossings}, with $u$, $B_c$ and exponent $p=1/\nu \psi +\omega / \psi$ as adjustable parameters.  $T_0= \unit[1.21]{K}$ was fixed as determined from the fit in Fig.~\ref{fig:activated}.  Best fit was achieved for $u=-5.56 \times 10^{-3}, p=2.40$, and $B_c=\unit[7.24]{T}$. }
\label{fig:xing_corrections}
\end{figure}

It should be noted that the use of Eq.~\ref{eq:activated} to collapse the data ignores the corrections to scaling, which are essential to the temperature-dependence of the crossing field.  These corrections vanish in the zero temperature limit, where the crossing fields converge to a fixed value.  In the appendix we show that the corrections give rise to a shift in the crossing points $B_x(T)$.  This shift $\delta_x (T) = ( B_c -B_x(T))/B_c$ will take the form
\begin{equation}
\delta_x (T) \sim  u  \left( \ln{ \frac{T_0}{T} } \right) ^{-\frac{1}{\nu \psi}- \frac{\omega}{ \psi}  },
\label{eq:delta_crossings}
\end{equation}
where $u$ is the leading irrelevant variable responsible for the corrections, and $\omega$ is the associated exponent.  The crossing fields shown in the inset to Fig.~\ref{fig:RvsH} are plotted as a function of temperature in Fig.~\ref{fig:xing_corrections}.  They are shown with a fit to Eq.~\ref{eq:delta_crossings}.  In the $T \rightarrow 0$, $\delta_x \rightarrow 0$ and the crossing fields approach $B_c$.  The zero-temperature limit of the crossing fields in Fig.~\ref{fig:xing_corrections} is slightly higher, but within 0.3\%, of the $B_c$ used for best collapse of the data shown in Fig.~\ref{fig:activated}.

\section{Discussion and conclusion}
\label{Sec:Con}

Let us first comment on the sources of disorder in the InO\textsubscript {x} films used here~\cite{OVAD1}. As an amorphous material, disorder occurs on the atomic length scale, based on randomness in interatomic spacings.  The compound also has between 5 and 30\% oxygen vacancies, which determine the carrier concentration. To preserve neutrality, some In atoms have a valence of +1 instead of +3.  This results in a random distribution of valence and charge fluctuations---a distribution which is thought to give rise to a stochiometric disorder~\cite{Givan} and may give rise to extended defect states~\cite{MatSci}. In addition to structural and chemical disorder, there is longer-scale disorder stemming from the films' characteristic undulating morphology.

Film characteristics depend on the interplay between the carrier concentration and the quenched disorder.  The former is largely fixed during deposition.  However the annealing process has the potential to drive a film from a highly disordered as-prepared nearly insulating state to a less disordered, and more metallic state~\cite{OVAD}.  Films in the lower mobility, highly disordered as-prepared state are known to exhibit direct quantum superconductor-insulator transitions, which can be analyzed using conventional power-law scaling. On the other hand, annealed films of higher mobilities have been shown here to exhibit a quantum superconductor-metal phase transition governed by an infinite-randomness critical point.

The explanation for this difference lies in the dynamics of rare, locally ordered regions close to the quantum phase transition. For the films in question, these rare regions are superconducting ``puddles'' immersed in an insulating or metallic matrix. According to the classification put forward in Refs.~\cite{VojtaSchmalian05,Vojta2014}, the rare region dimensionality needs to be at the lower critical dimension $d_c^-$ of the problem to produce quantum Griffiths singularities. Rare superconducting regions immersed in an insulating matrix are below $d_c^-$ and thus produce only exponentially small corrections to the conventional bulk critical behavior. In contrast, rare regions embedded in a metallic matrix are right at $d_c^-$  because the coupling to gapless electronic excitations causes Ohmic dissipation that slows down their dynamics \cite{VojtaSchmalian05}. Thus, a disorder tuned superconductor-metal transition is expected to feature quantum Griffiths singularities.

Note that these arguments require that the electrons which cause the dissipation can penetrate the entire ``puddle". Spivak et al.~\cite{Spivak, failed} pointed out that in the limit of large rare region size, the dissipation will scale with the surface of the rare region rather than its volume, cutting off the quantum Griffiths physics at the lowest temperatures. However, because of the exponential dependence of the rare-region energy scale on its size, this crossover temperature is expected to be extremely low, leaving a wide temperature regime governed by quantum Griffiths physics (see also Millis, Morr, Schmalian)\cite{Millis}.

The important question is under which conditions these quantum Griffiths singularities lead to activated scaling with a divergent dynamical critical exponent. This depends on whether the Harris criterion \cite{Harris} is satisfied or not. If the transition in the absence of disorder fulfills the Harris criterion, $d\nu  >2$, then even if Griffiths singularities exist, the dynamical exponent would not be expected to diverge. Alternatively, if the clean correlation exponent violates the Harris criterion, $z$ may diverge upon the introduction of quenched disorder \cite{Vojta2014}. In the case of a clean superconductor-metal transition tuned by magnetic field, $\nu  =\frac{1}{2}$ and $d =2$.  As a consequence, the Harris criterion is violated and $z$ is expected to diverge~\cite{Sachd}.

These general scaling arguments are confirmed by explicit model calculations. Hoyos et al. investigated the effects of dissipation on a disordered quantum phase transition with $O(N)$ order-parameter symmetry through the use of a strong-disorder renormalization group theory applied to the Landau-Ginzburg-Wilson field theory appropriate to the problem~\cite{Hoyos2007, Vojta2009}. They found that for Ohmic dissipation, the quantum phase transition is controlled by an infinite-randomness fixed point in the universality class of the random transverse-field Ising model. The dynamical scaling between the characteristic length scale $\xi$ and the corresponding time scale $\xi_\tau$ is not of power-law type, $\xi_\tau \sim \xi^z$, but activated, $\xi _{\tau } =\exp \left (const \times \xi ^{\psi }\right ) $, leading to Eq.~\ref{eq:activated}.

It is our conjecture that as the mobility of the films increases, and the high field state becomes metallic, rather than insulating, Ohmic dissipation increases and the quantum critical point changes from that of a conventional superconductor-insulator transition to an infinite-randomness critical point.


In summary, the quantum superconductor-metal transition of high-mobility amorphous InO\textsubscript {x} films tuned by perpendicular magnetic field exhibits quantum Griffiths effects which lead to an infinite-randomness quantum critical point. This is expected for systems with quenched disorder in the presence of Ohmic dissipation, and is  caused by the formation of large rare regions which are locally ordered superconducting puddles.

\begin{acknowledgments}
The authors would like to thank Rafael Fernandes and Steven Kivelson for helpful discussions. The work at Minnesota was supported by the National Science Foundation under Grant Nos. DMR-1209578 and DMR-1704456. Portions of this work were conducted  in the Minnesota Nano Center, which is supported by the National Science Foundation through the National Nano Coordinated Infrastructure Network (NNCI) under Award Number ECCS-1542202. T.V. acknowledges support by the NSF under Grant Nos. DMR-1506152, PHY-1125915 and PHY-1607611 as well as the hospitality of the Kavli Institute for Theoretical Physics, Santa Barbara, and the Aspen Center for Physics where parts of the work were performed.
\end{acknowledgments}

\section*{Appendix}

In this appendix we provide a derivation of Eq.~\ref{eq:NuZ_eff}, and a discussion of the effect of corrections to scaling. 
For simplicity, we first consider the case without corrections to scaling. Including such corrections is straight forward and does not change the results qualitatively. The standard power-law scaling form of the resistance at a magnetic-field-tuned transition is given by
\begin{equation}
R \left(\delta ,T \right) =\Phi \left (\delta T^{ -\frac{1}{\nu z}}\right)
\end{equation}
where $\Phi $ is the scaling function, $\nu $ is the correlation length exponent, $z$ is the dynamical critical exponent, and $\delta  =\left \vert  B -B_{c}\right \vert /B_{c}$ measures the distance from the critical field $B_{c}$. The critical resistance, $R_{c}=R (0 ,T) =\Phi (0)$. Curves of $R$ vs.\ $\delta$ at different temperatures cross at $\delta =0$. Now consider the slope of the resistance w.r.t. $\delta$,
\begin{equation}
S (T) =\left. \frac{ \partial R}{ \partial \delta } \right\vert _{\delta  =0} =T^{-\nicefrac{1}{\nu z}} \ \Phi ^{ \prime } (0)
\end{equation}from which it follows that
\begin{equation}
\frac{1}{\nu  z} = -\frac{d \ln  S}{d \ln  T}
\end{equation}
This value will be independent of $T$ as long as there are no corrections to scaling.

We now turn to activated scaling as expected for the random transverse field Ising model. The scaling form of the resistance is given by
\begin{equation}
R\left (\delta  , \ln{ \frac{T_{0}}{T}} \right  ) =\Phi  \left [\delta  \left (\ln  \frac{T_{0}}{T}\right )^{\frac{1}{\nu  \psi }}\right ]
\label{eq:activated2}
\end{equation}
Here, $\psi $ is the tunneling exponent. Note that at $\delta = 0$, this form predicts a single-valued critical resistance $R_c$, i.e. a single crossing point for isotherms. To find the exponent product $\nu\psi$, one can repeat the procedure used for power-law scaling,
\begin{equation}
S (T) = \left. \frac{ \partial R}{ \partial \delta }  \right\rvert_{\delta  =0} =\left (\ln  \frac{T_{0}}{T}\right )^{\frac{1}{\nu  \psi }} \phi ^{ \prime } (0)
\label{eq:S_act}
\end{equation}
Thus, $S(T)$ behaves as a power of $\ln {(T_0 /T)}$. Taking the appropriate logarithmic derivative yields
\begin{equation}
\frac{1}{\nu  \psi } =\frac{d \ln  S}{d\ln[ \ln {(T_0 /T)]}}.
\end{equation}
If there are no corrections to scaling this value will not depend upon temperature.
An extra complication stems from the microscopic scale, $T_{0}$, which is an additional fitting parameter.

Let us now work out what happens if one insists on analyzing data that follow the activated scaling form by using the procedure for power-law scaling. In other words, we calculate the logarithmic derivative w.r.t. $T$ of the slope given in Eq.~\ref{eq:S_act}:
\begin{eqnarray}
\left( \frac{1}{\nu  z} \right)_{\rm eff} &=& -\frac{d \ln{S}}{d \ln{T}}\\
    &=& -\frac{d}{d \ln  T} \ln{ \left [\phi ^{ \prime } \left (0\right ) \left (\ln  \frac{T_{0}}{T}\right )^{\frac{1}{\nu  \psi }}\right ]} \\
    &=& -\frac{d}{d \ln  T} \left [\frac{1}{\nu  \psi } \ln  \left (\ln  \frac{T_{0}}{T}\right )\right ].
\end{eqnarray}
This becomes
\begin{equation}
\begin{aligned}
\left( \frac{1}{\nu  z} \right)_{\rm eff} =& -\frac{1}{\nu  \psi } \, \frac{1}{\ln {(T_0 /T)}} \frac{d \ln {(T_0 /T)}}{d \ln  T} \\
=&\frac{1}{\nu  \psi }\, \frac{1}{\ln {(T_0 /T)}},
\end{aligned}
\label{eq:nuzeff}
\end{equation}
which completes the derivation of Eq.~\ref{eq:NuZ_eff}.

The r.h.s. of the last expression vanishes in the limit of zero temperature, which means that $\nu  z$ determined in this manner would diverge in the zero-temperature limit.

A similar result is obtained if corrections to scaling are included in the derivation, where the corrections are due to a leading irrelevant variable, $u$. Near an infinite randomness critical point the inverse disorder strength serves as an irrelevant scaling variable.  The form then becomes
\begin{equation}
R \left(  \delta , \ln{ \frac{T_0}{T} } , u \right) =
\bar{\Phi} \left[ \delta \left(   \ln{ \frac{T_0}{T} } \right) ^{\frac{1}{\nu \psi}},  u \left( \ln{ \frac{T_0}{T} } \right) ^{- \frac{\omega}{ \psi}  } \right]
\label{eq:add_corrections}
\end{equation}
where the exponent $\omega > 0$.  We expand the scaling function in the second argument, such that
\begin{equation}
\begin{aligned}
R \left(  \delta , \ln{ \frac{T_0}{T} } , u \right) & =
 \Phi \left[ \delta \left(   \ln{ \frac{T_0}{T} } \right) ^{\frac{1}{\nu \psi}} \right] \\
 +  & u  \left( \ln{ \frac{T_0}{T} } \right) ^{- \frac{\omega}{ \psi}  }
 \Phi_u \left[ \delta \left(   \ln{ \frac{T_0}{T} } \right) ^{\frac{1}{\nu \psi}} \right]
\end{aligned}
\label{eq:expanded_w_corrections}
\end{equation}
where both $\Phi$ and $\Phi_u$ are unknown functions.

One consequence of this corrected scaling form is that $R$ is not temperature-independent at $\delta = 0$, as in Eq.~\ref{eq:activated2}.  Instead, $R (0, \ln{ T_0 /T}, u ) = R_c + u ( \ln{ T_0/T})^{-\nicefrac{\omega }{\psi}} \Psi_n (0)$, where $R$ only approaches $R_c$ as $T \rightarrow 0$.  A second consequence is that $R(\delta)$ curves at finite temperatures do not cross right at $\delta = 0$.  The crossing points shift as a function of temperature, and approach $\delta = 0$ in the $T \rightarrow 0$ limit.

The temperature-dependence of the crossing points, $B_x(T)$, can be determined by expanding the scaling functions in Eq.~\ref{eq:expanded_w_corrections} linearly around $\delta = 0$ and determining where two isotherms cross. Let us take isotherms at $T$ and $2T$, though any multiplier can be used, and determine where
\begin{equation}
\begin{aligned}
R_c + \delta \left( \ln{ \frac{T_0}{T}} \right) ^{\frac{1}{\nu \psi} } \Phi ^{\prime} (0) + &u  \left( \ln{ \frac{T_0}{T} } \right) ^{- \frac{\omega}{ \psi}}
 \Phi_u (0) \\
 = R_c + \delta \left( \ln{ \frac{T_0}{2T} } \right) ^{\frac{1}{\nu \psi} } \Phi ^{\prime} (0) + &u  \left( \ln{ \frac{T_0}{2T} } \right) ^{- \frac{\omega}{ \psi} }
 \Phi_u (0).
 \end{aligned}
 \label{eq:crossing_condition}
\end{equation}
Critical resistance $R_c$ here is the value at the critical point without corrections, as in Eq.~\ref{eq:activated2}.  This equation can be rearranged and simplified by making the approximation
\begin{equation}
\begin{aligned}
\left( \ln{ \frac{T_0}{2T}  } \right) ^{\frac{1}{\nu \psi} }  =  &\left( \ln{ \frac{T_0}{T}  } - \ln{2} \right) ^{\frac{1}{\nu \psi} } \\
= & \left( \ln{ \frac{T_0}{2T}  } \right) ^{\frac{1}{\nu \psi} }      \left(1-  \frac{ \ln{2}}{\ln{T_0/T}}   \right) ^{\frac{1}{\nu \psi} } \\
\approx & \left( \ln{ \frac{T_0}{2T}  } \right) ^{\frac{1}{\nu \psi} }  \left( 1-  \frac{ \ln{2}}{\nu \psi} \frac{1}{\ln{T_0/T}}   \right)  .
\end{aligned}
\end{equation}
This can be used to show that the difference between the scaling terms can be written
\begin{equation}
\left( \ln{ \frac{T_0}{T} } \right) ^{- \frac{1}{ \nu \psi} }  -  \left( \ln{ \frac{T_0}{2T} } \right) ^{- \frac{1}{\nu \psi}}
= - \frac{\omega \ln{2}}{\psi}  \left( \ln{ \frac{T_0}{T} } \right) ^{- \frac{1}{\nu \psi} -1 },
\end{equation}
while, analogously, the correction terms can be written
\begin{equation}
\left( \ln{ \frac{T_0}{T} } \right) ^{- \frac{\omega}{ \psi} }  -  \left( \ln{ \frac{T_0}{2T} } \right) ^{- \frac{\omega}{ \psi}}
= - \frac{\omega \ln{2}}{\psi}  \left( \ln{ \frac{T_0}{T} } \right) ^{- \frac{\omega}{ \psi} -1 }.
\end{equation}
This can be inserted into the crossing condition, Eq.~\ref{eq:crossing_condition}, to show that the crossing points vary with temperature as
\begin{equation}
\delta_x (T) \sim  u  \left( \ln{ \frac{T_0}{T} } \right) ^{-\frac{1}{\nu \psi}- \frac{\omega}{ \psi}  }
\end{equation}
where $ \delta_x (T) = ((B_c - B_x (T))/B_c $ and $B_x(T)$ is the crossing point.  Deviation of the crossing point $\delta_x$ vanishes as $T \rightarrow 0$.

By following a calculation similar to that leading to Eq.\ \ref{eq:nuzeff},
it can also be shown that if the data are analyzed using power law scaling at the crossing points,
\begin{equation}
\left( \frac{1}{\nu z} \right) _{\rm eff} =\left( \frac{1}{\nu \psi } \right) _{\rm eff} \frac{1}{\ln ({T_{0}}/{T})},
\end{equation}
where
\begin{equation}
\left( \frac{1}{\nu \psi } \right) _{\rm eff} = \frac{1}{\nu \psi} - \frac{a \omega}{\psi} \left( \ln{ \frac{T_0 }{ T } } \right) ^{- \frac{ \omega}{\psi} },
\end{equation}
where $a = (c \Phi^{\prime \prime}_u (0) + u \Phi^{\prime}_u (0))/ \Phi ^{\prime}(0)$.  In the limit of $T \rightarrow 0$,  $(1/ \nu z) _{\rm eff} \rightarrow 0$ and  $(1/ \nu \psi) _{\rm eff} \rightarrow 1/ \nu \psi$ .

\bibliographystyle{apsrev4-1}
\bibliography{RareRegions}

\begin{thebibliography}{45}%
\makeatletter
\providecommand \@ifxundefined [1]{%
 \@ifx{#1\undefined}
}%
\providecommand \@ifnum [1]{%
 \ifnum #1\expandafter \@firstoftwo
 \else \expandafter \@secondoftwo
 \fi
}%
\providecommand \@ifx [1]{%
 \ifx #1\expandafter \@firstoftwo
 \else \expandafter \@secondoftwo
 \fi
}%
\providecommand \natexlab [1]{#1}%
\providecommand \enquote  [1]{``#1''}%
\providecommand \bibnamefont  [1]{#1}%
\providecommand \bibfnamefont [1]{#1}%
\providecommand \citenamefont [1]{#1}%
\providecommand \href@noop [0]{\@secondoftwo}%
\providecommand \href [0]{\begingroup \@sanitize@url \@href}%
\providecommand \@href[1]{\@@startlink{#1}\@@href}%
\providecommand \@@href[1]{\endgroup#1\@@endlink}%
\providecommand \@sanitize@url [0]{\catcode `\\12\catcode `\$12\catcode
  `\&12\catcode `\#12\catcode `\^12\catcode `\_12\catcode `\%12\relax}%
\providecommand \@@startlink[1]{}%
\providecommand \@@endlink[0]{}%
\providecommand \url  [0]{\begingroup\@sanitize@url \@url }%
\providecommand \@url [1]{\endgroup\@href {#1}{\urlprefix }}%
\providecommand \urlprefix  [0]{URL }%
\providecommand \Eprint [0]{\href }%
\providecommand \doibase [0]{http://dx.doi.org/}%
\providecommand \selectlanguage [0]{\@gobble}%
\providecommand \bibinfo  [0]{\@secondoftwo}%
\providecommand \bibfield  [0]{\@secondoftwo}%
\providecommand \translation [1]{[#1]}%
\providecommand \BibitemOpen [0]{}%
\providecommand \bibitemStop [0]{}%
\providecommand \bibitemNoStop [0]{.\EOS\space}%
\providecommand \EOS [0]{\spacefactor3000\relax}%
\providecommand \BibitemShut  [1]{\csname bibitem#1\endcsname}%
\let\auto@bib@innerbib\@empty
\bibitem [{\citenamefont {Hebard}\ and\ \citenamefont
  {Paalanen}(1990)}]{HebPal}%
  \BibitemOpen
  \bibfield  {author} {\bibinfo {author} {\bibfnamefont {A.~F.}\ \bibnamefont
  {Hebard}}\ and\ \bibinfo {author} {\bibfnamefont {M.~A.}\ \bibnamefont
  {Paalanen}},\ }\href@noop {} {\bibfield  {journal} {\bibinfo  {journal}
  {Phys. Rev. Lett.}\ }\textbf {\bibinfo {volume} {65}},\ \bibinfo {pages}
  {927} (\bibinfo {year} {1990})}\BibitemShut {NoStop}%
\bibitem [{\citenamefont {Yazdani}\ and\ \citenamefont
  {Kapitulnik}(1995)}]{Yaz}%
  \BibitemOpen
  \bibfield  {author} {\bibinfo {author} {\bibfnamefont {A.}~\bibnamefont
  {Yazdani}}\ and\ \bibinfo {author} {\bibfnamefont {A.}~\bibnamefont
  {Kapitulnik}},\ }\href@noop {} {\bibfield  {journal} {\bibinfo  {journal}
  {Phys. Rev. Lett.}\ }\textbf {\bibinfo {volume} {74}},\ \bibinfo {pages}
  {3037} (\bibinfo {year} {1995})}\BibitemShut {NoStop}%
\bibitem [{\citenamefont {Lin}\ \emph {et~al.}(2015)\citenamefont {Lin},
  \citenamefont {Nelson},\ and\ \citenamefont {Goldman}}]{Lin}%
  \BibitemOpen
  \bibfield  {author} {\bibinfo {author} {\bibfnamefont {Y.-H.}\ \bibnamefont
  {Lin}}, \bibinfo {author} {\bibfnamefont {J.}~\bibnamefont {Nelson}}, \ and\
  \bibinfo {author} {\bibfnamefont {A.~M.}\ \bibnamefont {Goldman}},\
  }\href@noop {} {\bibfield  {journal} {\bibinfo  {journal} {Physica
  C-Superconductivity and its Applications}\ }\textbf {\bibinfo {volume}
  {154}},\ \bibinfo {pages} {130} (\bibinfo {year} {2015})}\BibitemShut
  {NoStop}%
\bibitem [{\citenamefont {Fisher}(1990)}]{FisherM}%
  \BibitemOpen
  \bibfield  {author} {\bibinfo {author} {\bibfnamefont {M.~P.~A.}\
  \bibnamefont {Fisher}},\ }\href@noop {} {\bibfield  {journal} {\bibinfo
  {journal} {Phys. Rev. Lett.}\ }\textbf {\bibinfo {volume} {65}},\ \bibinfo
  {pages} {923} (\bibinfo {year} {1990})}\BibitemShut {NoStop}%
\bibitem [{\citenamefont {Qin}\ \emph {et~al.}(2006)\citenamefont {Qin},
  \citenamefont {Vincente},\ and\ \citenamefont {Yoon}}]{Qin}%
  \BibitemOpen
  \bibfield  {author} {\bibinfo {author} {\bibfnamefont {Y.}~\bibnamefont
  {Qin}}, \bibinfo {author} {\bibfnamefont {C.~L.}\ \bibnamefont {Vincente}}, \
  and\ \bibinfo {author} {\bibfnamefont {J.}~\bibnamefont {Yoon}},\ }\href@noop
  {} {\bibfield  {journal} {\bibinfo  {journal} {Phys. Rev. B}\ }\textbf
  {\bibinfo {volume} {73}},\ \bibinfo {pages} {100505} (\bibinfo {year}
  {2006})}\BibitemShut {NoStop}%
\bibitem [{\citenamefont {Mason}\ and\ \citenamefont
  {Kapitulnik}(2001)}]{Mason}%
  \BibitemOpen
  \bibfield  {author} {\bibinfo {author} {\bibfnamefont {N.}~\bibnamefont
  {Mason}}\ and\ \bibinfo {author} {\bibfnamefont {A.}~\bibnamefont
  {Kapitulnik}},\ }\href@noop {} {\bibfield  {journal} {\bibinfo  {journal}
  {Phys. Rev. B}\ }\textbf {\bibinfo {volume} {64}},\ \bibinfo {pages} {00504}
  (\bibinfo {year} {2001})}\BibitemShut {NoStop}%
\bibitem [{\citenamefont {Kapitulnik}\ \emph {et~al.}()\citenamefont
  {Kapitulnik}, \citenamefont {Kivelson},\ and\ \citenamefont
  {Spivak}}]{failed}%
  \BibitemOpen
  \bibfield  {author} {\bibinfo {author} {\bibfnamefont {A.}~\bibnamefont
  {Kapitulnik}}, \bibinfo {author} {\bibfnamefont {S.~A.}\ \bibnamefont
  {Kivelson}}, \ and\ \bibinfo {author} {\bibfnamefont {B.}~\bibnamefont
  {Spivak}},\ }\href@noop {} {\bibinfo  {journal} {arXiv:1712.07215v1
  [cond-mat.supr-con]}\ }\BibitemShut {NoStop}%
\bibitem [{\citenamefont {Tsen}\ \emph {et~al.}(2016)\citenamefont {Tsen},
  \citenamefont {Hunt}, \citenamefont {Kim}, \citenamefont {Yuan},
  \citenamefont {Jia}, \citenamefont {Cava}, \citenamefont {Hone},
  \citenamefont {Kim}, \citenamefont {Dean},\ and\ \citenamefont
  {Pasupathy}}]{Tsen}%
  \BibitemOpen
\bibfield  {journal} {  }\bibfield  {author} {\bibinfo {author} {\bibfnamefont
  {A.~W.}\ \bibnamefont {Tsen}}, \bibinfo {author} {\bibfnamefont
  {B.}~\bibnamefont {Hunt}}, \bibinfo {author} {\bibfnamefont {Y.~D.}\
  \bibnamefont {Kim}}, \bibinfo {author} {\bibfnamefont {Z.~J.}\ \bibnamefont
  {Yuan}}, \bibinfo {author} {\bibfnamefont {S.}~\bibnamefont {Jia}}, \bibinfo
  {author} {\bibfnamefont {R.~J.}\ \bibnamefont {Cava}}, \bibinfo {author}
  {\bibfnamefont {J.}~\bibnamefont {Hone}}, \bibinfo {author} {\bibfnamefont
  {P.}~\bibnamefont {Kim}}, \bibinfo {author} {\bibfnamefont {C.~R.}\
  \bibnamefont {Dean}}, \ and\ \bibinfo {author} {\bibfnamefont {A.~N.}\
  \bibnamefont {Pasupathy}},\ }\href@noop {} {\bibfield  {journal} {\bibinfo
  {journal} {Nat. Phys.}\ }\textbf {\bibinfo {volume} {12}},\ \bibinfo {pages}
  {208} (\bibinfo {year} {2016})}\BibitemShut {NoStop}%
\bibitem [{\citenamefont {Das}\ and\ \citenamefont {Doniach}(1999)}]{Das1}%
  \BibitemOpen
  \bibfield  {author} {\bibinfo {author} {\bibfnamefont {D.}~\bibnamefont
  {Das}}\ and\ \bibinfo {author} {\bibfnamefont {S.}~\bibnamefont {Doniach}},\
  }\href@noop {} {\bibfield  {journal} {\bibinfo  {journal} {Phys. Rev. B}\
  }\textbf {\bibinfo {volume} {60}},\ \bibinfo {pages} {1261} (\bibinfo {year}
  {1999})}\BibitemShut {NoStop}%
\bibitem [{\citenamefont {Das}\ and\ \citenamefont {Doniach}(2001)}]{Das2}%
  \BibitemOpen
  \bibfield  {author} {\bibinfo {author} {\bibfnamefont {D.}~\bibnamefont
  {Das}}\ and\ \bibinfo {author} {\bibfnamefont {S.}~\bibnamefont {Doniach}},\
  }\href@noop {} {\bibfield  {journal} {\bibinfo  {journal} {Phys. Rev. B}\
  }\textbf {\bibinfo {volume} {64}},\ \bibinfo {pages} {134511} (\bibinfo
  {year} {2001})}\BibitemShut {NoStop}%
\bibitem [{\citenamefont {Dalidovich}\ and\ \citenamefont
  {Phillips}(2002)}]{Dalid1}%
  \BibitemOpen
  \bibfield  {author} {\bibinfo {author} {\bibfnamefont {D.}~\bibnamefont
  {Dalidovich}}\ and\ \bibinfo {author} {\bibfnamefont {P.}~\bibnamefont
  {Phillips}},\ }\href@noop {} {\bibfield  {journal} {\bibinfo  {journal}
  {Phys. Rev. Lett.}\ }\textbf {\bibinfo {volume} {89}},\ \bibinfo {pages}
  {027001} (\bibinfo {year} {2002})}\BibitemShut {NoStop}%
\bibitem [{\citenamefont {Phillips}\ and\ \citenamefont
  {Dalidovich}(2003)}]{Dalid2}%
  \BibitemOpen
  \bibfield  {author} {\bibinfo {author} {\bibfnamefont {P.}~\bibnamefont
  {Phillips}}\ and\ \bibinfo {author} {\bibfnamefont {D.}~\bibnamefont
  {Dalidovich}},\ }\href@noop {} {\bibfield  {journal} {\bibinfo  {journal}
  {Science}\ }\textbf {\bibinfo {volume} {302}},\ \bibinfo {pages} {243}
  (\bibinfo {year} {2003})}\BibitemShut {NoStop}%
\bibitem [{\citenamefont {Tamir}\ \emph {et~al.}(2018)\citenamefont {Tamir},
  \citenamefont {Benyamini}, \citenamefont {Telford}, \citenamefont
  {Gorniaczyk}, \citenamefont {Doron}, \citenamefont {Levinson}, \citenamefont
  {Wang}, \citenamefont {Gay}, \citenamefont {Sac\'{e}p\'{e}}, \citenamefont
  {Hone}, \citenamefont {Watanabe}, \citenamefont {Taniguchi}, \citenamefont
  {Dean}, \citenamefont {Pasupathy},\ and\ \citenamefont {Shahar}}]{Tamir2018}%
  \BibitemOpen
  \bibfield  {author} {\bibinfo {author} {\bibfnamefont {I.}~\bibnamefont
  {Tamir}}, \bibinfo {author} {\bibfnamefont {A.}~\bibnamefont {Benyamini}},
  \bibinfo {author} {\bibfnamefont {E.~J.}\ \bibnamefont {Telford}}, \bibinfo
  {author} {\bibfnamefont {F.}~\bibnamefont {Gorniaczyk}}, \bibinfo {author}
  {\bibfnamefont {A.}~\bibnamefont {Doron}}, \bibinfo {author} {\bibfnamefont
  {T.}~\bibnamefont {Levinson}}, \bibinfo {author} {\bibfnamefont
  {D.}~\bibnamefont {Wang}}, \bibinfo {author} {\bibfnamefont {F.}~\bibnamefont
  {Gay}}, \bibinfo {author} {\bibfnamefont {B.}~\bibnamefont {Sac\'{e}p\'{e}}},
  \bibinfo {author} {\bibfnamefont {J.}~\bibnamefont {Hone}}, \bibinfo {author}
  {\bibfnamefont {K.}~\bibnamefont {Watanabe}}, \bibinfo {author}
  {\bibfnamefont {T.}~\bibnamefont {Taniguchi}}, \bibinfo {author}
  {\bibfnamefont {C.~R.}\ \bibnamefont {Dean}}, \bibinfo {author}
  {\bibfnamefont {A.~N.}\ \bibnamefont {Pasupathy}}, \ and\ \bibinfo {author}
  {\bibfnamefont {D.}~\bibnamefont {Shahar}},\ }\href@noop {} {\bibfield
  {journal} {\bibinfo  {journal} {arXiv:1804.04648v1 [cond-mat.supr-con]}\ }
  (\bibinfo {year} {2018})}\BibitemShut {NoStop}%
\bibitem [{\citenamefont {Sondhi}\ \emph {et~al.}(1997)\citenamefont {Sondhi},
  \citenamefont {Girvin}, \citenamefont {Carini},\ and\ \citenamefont
  {Shahar}}]{SGCS}%
  \BibitemOpen
  \bibfield  {author} {\bibinfo {author} {\bibfnamefont {S.~L.}\ \bibnamefont
  {Sondhi}}, \bibinfo {author} {\bibfnamefont {S.~M.}\ \bibnamefont {Girvin}},
  \bibinfo {author} {\bibfnamefont {J.~P.}\ \bibnamefont {Carini}}, \ and\
  \bibinfo {author} {\bibfnamefont {D.}~\bibnamefont {Shahar}},\ }\href@noop {}
  {\bibfield  {journal} {\bibinfo  {journal} {Rev. Mod. Phys.}\ }\textbf
  {\bibinfo {volume} {69}},\ \bibinfo {pages} {315} (\bibinfo {year}
  {1997})}\BibitemShut {NoStop}%
\bibitem [{\citenamefont {Gantmakher}\ \emph {et~al.}(2000)\citenamefont
  {Gantmakher}, \citenamefont {Golubkov}, \citenamefont {Dolgopolov},
  \citenamefont {Tsydynzhapov},\ and\ \citenamefont {Shashkin}}]{Gantm2000}%
  \BibitemOpen
  \bibfield  {author} {\bibinfo {author} {\bibfnamefont {V.~F.}\ \bibnamefont
  {Gantmakher}}, \bibinfo {author} {\bibfnamefont {M.~V.}\ \bibnamefont
  {Golubkov}}, \bibinfo {author} {\bibfnamefont {V.~T.}\ \bibnamefont
  {Dolgopolov}}, \bibinfo {author} {\bibfnamefont {G.~E.}\ \bibnamefont
  {Tsydynzhapov}}, \ and\ \bibinfo {author} {\bibfnamefont {A.~A.}\
  \bibnamefont {Shashkin}},\ }\href@noop {} {\bibfield  {journal} {\bibinfo
  {journal} {JETP Lett.}\ }\textbf {\bibinfo {volume} {71}},\ \bibinfo {pages}
  {160} (\bibinfo {year} {2000})}\BibitemShut {NoStop}%
\bibitem [{\citenamefont {Xing}\ \emph {et~al.}(2015)\citenamefont {Xing},
  \citenamefont {Zhang}, \citenamefont {Fu}, \citenamefont {Liu}, \citenamefont
  {Sun}, \citenamefont {Peng}, \citenamefont {Wang}, \citenamefont {Lin},
  \citenamefont {Ma}, \citenamefont {Xue}, \citenamefont {Wang},\ and\
  \citenamefont {Xie}}]{xing}%
  \BibitemOpen
  \bibfield  {author} {\bibinfo {author} {\bibfnamefont {Y.}~\bibnamefont
  {Xing}}, \bibinfo {author} {\bibfnamefont {H.-M.}\ \bibnamefont {Zhang}},
  \bibinfo {author} {\bibfnamefont {H.-L.}\ \bibnamefont {Fu}}, \bibinfo
  {author} {\bibfnamefont {H.}~\bibnamefont {Liu}}, \bibinfo {author}
  {\bibfnamefont {Y.}~\bibnamefont {Sun}}, \bibinfo {author} {\bibfnamefont
  {J.-P.}\ \bibnamefont {Peng}}, \bibinfo {author} {\bibfnamefont
  {F.}~\bibnamefont {Wang}}, \bibinfo {author} {\bibfnamefont {X.}~\bibnamefont
  {Lin}}, \bibinfo {author} {\bibfnamefont {X.-C.}\ \bibnamefont {Ma}},
  \bibinfo {author} {\bibfnamefont {Q.-K.}\ \bibnamefont {Xue}}, \bibinfo
  {author} {\bibfnamefont {J.}~\bibnamefont {Wang}}, \ and\ \bibinfo {author}
  {\bibfnamefont {X.~C.}\ \bibnamefont {Xie}},\ }\href@noop {} {\bibfield
  {journal} {\bibinfo  {journal} {Science}\ }\textbf {\bibinfo {volume}
  {350}},\ \bibinfo {pages} {542} (\bibinfo {year} {2015})}\BibitemShut
  {NoStop}%
\bibitem [{\citenamefont {Shen}\ \emph {et~al.}(2016)\citenamefont {Shen},
  \citenamefont {Xing}, \citenamefont {Wang}, \citenamefont {Liu},
  \citenamefont {Fu}, \citenamefont {Zhang}, \citenamefont {He}, \citenamefont
  {Xie}, \citenamefont {Nie},\ and\ \citenamefont {Wang}}]{Shen}%
  \BibitemOpen
  \bibfield  {author} {\bibinfo {author} {\bibfnamefont {S.}~\bibnamefont
  {Shen}}, \bibinfo {author} {\bibfnamefont {Y.}~\bibnamefont {Xing}}, \bibinfo
  {author} {\bibfnamefont {P.}~\bibnamefont {Wang}}, \bibinfo {author}
  {\bibfnamefont {H.}~\bibnamefont {Liu}}, \bibinfo {author} {\bibfnamefont
  {H.-L.}\ \bibnamefont {Fu}}, \bibinfo {author} {\bibfnamefont
  {Y.}~\bibnamefont {Zhang}}, \bibinfo {author} {\bibfnamefont
  {L.}~\bibnamefont {He}}, \bibinfo {author} {\bibfnamefont {X.}~\bibnamefont
  {Xie}, \bibfnamefont {X.~C.and~Lin}}, \bibinfo {author} {\bibfnamefont
  {J.}~\bibnamefont {Nie}}, \ and\ \bibinfo {author} {\bibfnamefont
  {J.}~\bibnamefont {Wang}},\ }\href@noop {} {\bibfield  {journal} {\bibinfo
  {journal} {Phys. Rev. B}\ }\textbf {\bibinfo {volume} {94}},\ \bibinfo
  {pages} {144517} (\bibinfo {year} {2016})}\BibitemShut {NoStop}%
\bibitem [{\citenamefont {Saito}\ \emph {et~al.}(2018)\citenamefont {Saito},
  \citenamefont {Nojima},\ and\ \citenamefont {Iwasa}}]{Saito2018}%
  \BibitemOpen
  \bibfield  {author} {\bibinfo {author} {\bibfnamefont {Y.}~\bibnamefont
  {Saito}}, \bibinfo {author} {\bibfnamefont {T.}~\bibnamefont {Nojima}}, \
  and\ \bibinfo {author} {\bibfnamefont {Y.}~\bibnamefont {Iwasa}},\
  }\href@noop {} {\bibfield  {journal} {\bibinfo  {journal} {Nat. Comm.}\
  }\textbf {\bibinfo {volume} {9}},\ \bibinfo {pages} {778} (\bibinfo {year}
  {2018})}\BibitemShut {NoStop}%
\bibitem [{\citenamefont {Xing}\ \emph {et~al.}(2017)\citenamefont {Xing},
  \citenamefont {Zhao}, \citenamefont {Shan}, \citenamefont {Zheng},
  \citenamefont {Zhang}, \citenamefont {Fu}, \citenamefont {Liu}, \citenamefont
  {Tian}, \citenamefont {Xi}, \citenamefont {Liu}, \citenamefont {Feng},
  \citenamefont {Lin}, \citenamefont {Ji}, \citenamefont {Chen}, \citenamefont
  {Xue},\ and\ \citenamefont {Wang}}]{XingN}%
  \BibitemOpen
  \bibfield  {author} {\bibinfo {author} {\bibfnamefont {Y.}~\bibnamefont
  {Xing}}, \bibinfo {author} {\bibfnamefont {K.}~\bibnamefont {Zhao}}, \bibinfo
  {author} {\bibfnamefont {P.}~\bibnamefont {Shan}}, \bibinfo {author}
  {\bibfnamefont {F.}~\bibnamefont {Zheng}}, \bibinfo {author} {\bibfnamefont
  {Y.}~\bibnamefont {Zhang}}, \bibinfo {author} {\bibfnamefont
  {H.}~\bibnamefont {Fu}}, \bibinfo {author} {\bibfnamefont {Y.}~\bibnamefont
  {Liu}}, \bibinfo {author} {\bibfnamefont {M.}~\bibnamefont {Tian}}, \bibinfo
  {author} {\bibfnamefont {C.}~\bibnamefont {Xi}}, \bibinfo {author}
  {\bibfnamefont {H.}~\bibnamefont {Liu}}, \bibinfo {author} {\bibfnamefont
  {J.}~\bibnamefont {Feng}}, \bibinfo {author} {\bibfnamefont {X.}~\bibnamefont
  {Lin}}, \bibinfo {author} {\bibfnamefont {S.}~\bibnamefont {Ji}}, \bibinfo
  {author} {\bibfnamefont {X.}~\bibnamefont {Chen}}, \bibinfo {author}
  {\bibfnamefont {Q.-K.}\ \bibnamefont {Xue}}, \ and\ \bibinfo {author}
  {\bibfnamefont {J.}~\bibnamefont {Wang}},\ }\href@noop {} {\bibfield
  {journal} {\bibinfo  {journal} {Nano Lett.}\ }\textbf {\bibinfo {volume}
  {17}},\ \bibinfo {pages} {6802} (\bibinfo {year} {2017})}\BibitemShut
  {NoStop}%
\bibitem [{\citenamefont {Griffiths}(1969)}]{Grif}%
  \BibitemOpen
  \bibfield  {author} {\bibinfo {author} {\bibfnamefont {R.~B.}\ \bibnamefont
  {Griffiths}},\ }\href@noop {} {\bibfield  {journal} {\bibinfo  {journal}
  {Phys. Rev. Lett.}\ }\textbf {\bibinfo {volume} {23}},\ \bibinfo {pages} {17}
  (\bibinfo {year} {1969})}\BibitemShut {NoStop}%
\bibitem [{\citenamefont {Thill}\ and\ \citenamefont
  {Huse}(1995)}]{ThillHuse95}%
  \BibitemOpen
  \bibfield  {author} {\bibinfo {author} {\bibfnamefont {M.}~\bibnamefont
  {Thill}}\ and\ \bibinfo {author} {\bibfnamefont {D.~A.}\ \bibnamefont
  {Huse}},\ }\href {\doibase 10.1016/0378-4371(94)00247-Q} {\bibfield
  {journal} {\bibinfo  {journal} {Physica A}\ }\textbf {\bibinfo {volume}
  {214}},\ \bibinfo {pages} {321} (\bibinfo {year} {1995})}\BibitemShut
  {NoStop}%
\bibitem [{\citenamefont {Young}\ and\ \citenamefont
  {Rieger}(1996)}]{YoungRieger96}%
  \BibitemOpen
  \bibfield  {author} {\bibinfo {author} {\bibfnamefont {A.~P.}\ \bibnamefont
  {Young}}\ and\ \bibinfo {author} {\bibfnamefont {H.}~\bibnamefont {Rieger}},\
  }\href@noop {} {\bibfield  {journal} {\bibinfo  {journal} {Phys. Rev. B}\
  }\textbf {\bibinfo {volume} {53}},\ \bibinfo {pages} {8486} (\bibinfo {year}
  {1996})}\BibitemShut {NoStop}%
\bibitem [{\citenamefont {Fisher}(1995)}]{Fisher95}%
  \BibitemOpen
  \bibfield  {author} {\bibinfo {author} {\bibfnamefont {D.~S.}\ \bibnamefont
  {Fisher}},\ }\href {\doibase 10.1103/PhysRevB.51.6411} {\bibfield  {journal}
  {\bibinfo  {journal} {Phys. Rev. B}\ }\textbf {\bibinfo {volume} {51}},\
  \bibinfo {pages} {6411} (\bibinfo {year} {1995})}\BibitemShut {NoStop}%
\bibitem [{\citenamefont {Motrunich}\ \emph {et~al.}(2000)\citenamefont
  {Motrunich}, \citenamefont {Mau}, \citenamefont {Huse},\ and\ \citenamefont
  {Fisher}}]{Motrunich}%
  \BibitemOpen
  \bibfield  {author} {\bibinfo {author} {\bibfnamefont {O.}~\bibnamefont
  {Motrunich}}, \bibinfo {author} {\bibfnamefont {S.-C.}\ \bibnamefont {Mau}},
  \bibinfo {author} {\bibfnamefont {D.~A.}\ \bibnamefont {Huse}}, \ and\
  \bibinfo {author} {\bibfnamefont {D.~S.}\ \bibnamefont {Fisher}},\
  }\href@noop {} {\bibfield  {journal} {\bibinfo  {journal} {Phys. Rev. B}\
  }\textbf {\bibinfo {volume} {61}},\ \bibinfo {pages} {1160} (\bibinfo {year}
  {2000})}\BibitemShut {NoStop}%
\bibitem [{\citenamefont {Hoyos}\ \emph {et~al.}(2007)\citenamefont {Hoyos},
  \citenamefont {Kotabage},\ and\ \citenamefont {Vojta}}]{Hoyos2007}%
  \BibitemOpen
  \bibfield  {author} {\bibinfo {author} {\bibfnamefont {J.~A.}\ \bibnamefont
  {Hoyos}}, \bibinfo {author} {\bibfnamefont {C.}~\bibnamefont {Kotabage}}, \
  and\ \bibinfo {author} {\bibfnamefont {T.}~\bibnamefont {Vojta}},\
  }\href@noop {} {\bibfield  {journal} {\bibinfo  {journal} {Phys. Rev. Lett}\
  }\textbf {\bibinfo {volume} {99}},\ \bibinfo {pages} {260601} (\bibinfo
  {year} {2007})}\BibitemShut {NoStop}%
\bibitem [{\citenamefont {Vojta}\ \emph
  {et~al.}(2009{\natexlab{a}})\citenamefont {Vojta}, \citenamefont {Kotabage},\
  and\ \citenamefont {Hoyos}}]{Vojta2009}%
  \BibitemOpen
  \bibfield  {author} {\bibinfo {author} {\bibfnamefont {T.}~\bibnamefont
  {Vojta}}, \bibinfo {author} {\bibfnamefont {C.}~\bibnamefont {Kotabage}}, \
  and\ \bibinfo {author} {\bibfnamefont {J.~A.}\ \bibnamefont {Hoyos}},\
  }\href@noop {} {\bibfield  {journal} {\bibinfo  {journal} {Phys. Rev. B}\
  }\textbf {\bibinfo {volume} {79}},\ \bibinfo {pages} {024401} (\bibinfo
  {year} {2009}{\natexlab{a}})}\BibitemShut {NoStop}%
\bibitem [{\citenamefont {Vojta}(2006)}]{Vojta06}%
  \BibitemOpen
  \bibfield  {author} {\bibinfo {author} {\bibfnamefont {T.}~\bibnamefont
  {Vojta}},\ }\href {\doibase 10.1088/0305-4470/39/22/R01} {\bibfield
  {journal} {\bibinfo  {journal} {J. Phys. A}\ }\textbf {\bibinfo {volume}
  {39}},\ \bibinfo {pages} {R143} (\bibinfo {year} {2006})}\BibitemShut
  {NoStop}%
\bibitem [{\citenamefont {Vojta}(2010)}]{VojtaRev}%
  \BibitemOpen
  \bibfield  {author} {\bibinfo {author} {\bibfnamefont {T.}~\bibnamefont
  {Vojta}},\ }\href@noop {} {\bibfield  {journal} {\bibinfo  {journal} {J. Low
  Temp. Phys.}\ }\textbf {\bibinfo {volume} {164}},\ \bibinfo {pages} {299}
  (\bibinfo {year} {2010})}\BibitemShut {NoStop}%
\bibitem [{\citenamefont {Del~Maestro}\ \emph {et~al.}(2010)\citenamefont
  {Del~Maestro}, \citenamefont {Rosenow}, \citenamefont {Hoyos},\ and\
  \citenamefont {Vojta}}]{DRHV10}%
  \BibitemOpen
  \bibfield  {author} {\bibinfo {author} {\bibfnamefont {A.}~\bibnamefont
  {Del~Maestro}}, \bibinfo {author} {\bibfnamefont {B.}~\bibnamefont
  {Rosenow}}, \bibinfo {author} {\bibfnamefont {J.~A.}\ \bibnamefont {Hoyos}},
  \ and\ \bibinfo {author} {\bibfnamefont {T.}~\bibnamefont {Vojta}},\ }\href
  {\doibase 10.1103/PhysRevLett.105.145702} {\bibfield  {journal} {\bibinfo
  {journal} {Phys. Rev. Lett.}\ }\textbf {\bibinfo {volume} {105}},\ \bibinfo
  {pages} {145702} (\bibinfo {year} {2010})}\BibitemShut {NoStop}%
\bibitem [{\citenamefont {Ovadyahu}(1986)}]{OVAD}%
  \BibitemOpen
  \bibfield  {author} {\bibinfo {author} {\bibfnamefont {Z.}~\bibnamefont
  {Ovadyahu}},\ }\href@noop {} {\bibfield  {journal} {\bibinfo  {journal} {J.
  Phys. C: Solid State Physics}\ }\textbf {\bibinfo {volume} {19}},\ \bibinfo
  {pages} {5187} (\bibinfo {year} {1986})}\BibitemShut {NoStop}%
\bibitem [{\citenamefont {Altshuler}\ \emph {et~al.}(1982)\citenamefont
  {Altshuler}, \citenamefont {Aronov}, \citenamefont {Khmelnitskii},\ and\
  \citenamefont {Larkin}}]{Alt}%
  \BibitemOpen
  \bibfield  {author} {\bibinfo {author} {\bibfnamefont {B.~I.}\ \bibnamefont
  {Altshuler}}, \bibinfo {author} {\bibfnamefont {A.~G.}\ \bibnamefont
  {Aronov}}, \bibinfo {author} {\bibfnamefont {D.~E.}\ \bibnamefont
  {Khmelnitskii}}, \ and\ \bibinfo {author} {\bibfnamefont {A.~I.}\
  \bibnamefont {Larkin}},\ }\href@noop {} {\emph {\bibinfo {title} {Quantum
  Theory of Solids}}},\ edited by\ \bibinfo {editor} {\bibfnamefont {I.~M.}\
  \bibnamefont {Lifshits}},\ Physics Series\ (\bibinfo  {publisher} {MIR
  Publishers, Moscow},\ \bibinfo {year} {1982})\BibitemShut {NoStop}%
\bibitem [{\citenamefont {Kramer}\ and\ \citenamefont
  {MacKinnon}(1993)}]{Kramer}%
  \BibitemOpen
  \bibfield  {author} {\bibinfo {author} {\bibfnamefont {B.}~\bibnamefont
  {Kramer}}\ and\ \bibinfo {author} {\bibfnamefont {A.}~\bibnamefont
  {MacKinnon}},\ }\href@noop {} {\bibfield  {journal} {\bibinfo  {journal}
  {Rep. Prog. Phys.}\ }\textbf {\bibinfo {volume} {56}},\ \bibinfo {pages}
  {1469} (\bibinfo {year} {1993})}\BibitemShut {NoStop}%
\bibitem [{\citenamefont {Vojta}\ \emph
  {et~al.}(2009{\natexlab{b}})\citenamefont {Vojta}, \citenamefont {Farquhar},\
  and\ \citenamefont {Mast}}]{Vojta2009a}%
  \BibitemOpen
  \bibfield  {author} {\bibinfo {author} {\bibfnamefont {T.}~\bibnamefont
  {Vojta}}, \bibinfo {author} {\bibfnamefont {A.}~\bibnamefont {Farquhar}}, \
  and\ \bibinfo {author} {\bibfnamefont {J.}~\bibnamefont {Mast}},\ }\href@noop
  {} {\bibfield  {journal} {\bibinfo  {journal} {Phys. Rev. E}\ }\textbf
  {\bibinfo {volume} {79}},\ \bibinfo {pages} {011111} (\bibinfo {year}
  {2009}{\natexlab{b}})}\BibitemShut {NoStop}%
\bibitem [{\citenamefont {Kovacs}\ and\ \citenamefont
  {Igloi}(2010)}]{Kovac2010}%
  \BibitemOpen
  \bibfield  {author} {\bibinfo {author} {\bibfnamefont {I.~A.}\ \bibnamefont
  {Kovacs}}\ and\ \bibinfo {author} {\bibfnamefont {F.}~\bibnamefont {Igloi}},\
  }\href@noop {} {\bibfield  {journal} {\bibinfo  {journal} {Phys. Rev. B}\
  }\textbf {\bibinfo {volume} {82}},\ \bibinfo {pages} {054437} (\bibinfo
  {year} {2010})}\BibitemShut {NoStop}%
\bibitem [{\citenamefont {Del~Maestro}\ \emph {et~al.}(2008)\citenamefont
  {Del~Maestro}, \citenamefont {Rosenow}, \citenamefont {Muller},\ and\
  \citenamefont {Sachdev}}]{Maest2008}%
  \BibitemOpen
  \bibfield  {author} {\bibinfo {author} {\bibfnamefont {A.}~\bibnamefont
  {Del~Maestro}}, \bibinfo {author} {\bibfnamefont {B.}~\bibnamefont
  {Rosenow}}, \bibinfo {author} {\bibfnamefont {M.}~\bibnamefont {Muller}}, \
  and\ \bibinfo {author} {\bibfnamefont {S.}~\bibnamefont {Sachdev}},\
  }\href@noop {} {\bibfield  {journal} {\bibinfo  {journal} {Phys. Rev. Lett.}\
  }\textbf {\bibinfo {volume} {101}},\ \bibinfo {pages} {035701} (\bibinfo
  {year} {2008})}\BibitemShut {NoStop}%
\bibitem [{\citenamefont {Skinner}\ \emph {et~al.}()\citenamefont {Skinner},
  \citenamefont {Ruhman},\ and\ \citenamefont {Nahum}}]{Skinner2018}%
  \BibitemOpen
  \bibfield  {author} {\bibinfo {author} {\bibfnamefont {B.}~\bibnamefont
  {Skinner}}, \bibinfo {author} {\bibfnamefont {J.}~\bibnamefont {Ruhman}}, \
  and\ \bibinfo {author} {\bibfnamefont {A.}~\bibnamefont {Nahum}},\
  }\href@noop {} {\bibinfo  {journal} {arXiv:1808.05953v2
  [cond-mat.stat-mech]}\ }\BibitemShut {NoStop}%
\bibitem [{\citenamefont {Ovadyahu}(2017)}]{OVAD1}%
  \BibitemOpen
\bibfield  {journal} {  }\bibfield  {author} {\bibinfo {author} {\bibfnamefont
  {Z.}~\bibnamefont {Ovadyahu}},\ }\href@noop {} {\bibfield  {journal}
  {\bibinfo  {journal} {Phys. Rev. B}\ }\textbf {\bibinfo {volume} {95}},\
  \bibinfo {pages} {134203} (\bibinfo {year} {2017})}\BibitemShut {NoStop}%
\bibitem [{\citenamefont {Givan}\ and\ \citenamefont {Ovadyahu}(2012)}]{Givan}%
  \BibitemOpen
  \bibfield  {author} {\bibinfo {author} {\bibfnamefont {U.}~\bibnamefont
  {Givan}}\ and\ \bibinfo {author} {\bibfnamefont {Z.}~\bibnamefont
  {Ovadyahu}},\ }\href@noop {} {\bibfield  {journal} {\bibinfo  {journal}
  {Phys. Rev. B}\ }\textbf {\bibinfo {volume} {86}},\ \bibinfo {pages} {165101}
  (\bibinfo {year} {2012})}\BibitemShut {NoStop}%
\bibitem [{\citenamefont {Medvedeva}\ \emph {et~al.}()\citenamefont
  {Medvedeva}, \citenamefont {Buchholz},\ and\ \citenamefont {Chang}}]{MatSci}%
  \BibitemOpen
  \bibfield  {author} {\bibinfo {author} {\bibfnamefont {J.~E.}\ \bibnamefont
  {Medvedeva}}, \bibinfo {author} {\bibfnamefont {D.~B.}\ \bibnamefont
  {Buchholz}}, \ and\ \bibinfo {author} {\bibfnamefont {R.~P.~H.}\ \bibnamefont
  {Chang}},\ }\href@noop {} {\bibfield  {journal} {\bibinfo  {journal} {Adv.
  Electron. Mater.}\ }\textbf {\bibinfo {volume} {3}},\ \bibinfo {pages}
  {1700082}}\BibitemShut {NoStop}%
\bibitem [{\citenamefont {Vojta}\ and\ \citenamefont
  {Schmalian}(2005)}]{VojtaSchmalian05}%
  \BibitemOpen
  \bibfield  {author} {\bibinfo {author} {\bibfnamefont {T.}~\bibnamefont
  {Vojta}}\ and\ \bibinfo {author} {\bibfnamefont {J.}~\bibnamefont
  {Schmalian}},\ }\href {\doibase 10.1103/PhysRevB.72.045438} {\bibfield
  {journal} {\bibinfo  {journal} {Phys. Rev. B}\ }\textbf {\bibinfo {volume}
  {72}},\ \bibinfo {pages} {045438} (\bibinfo {year} {2005})}\BibitemShut
  {NoStop}%
\bibitem [{\citenamefont {Vojta}\ and\ \citenamefont
  {Hoyos}(2014)}]{Vojta2014}%
  \BibitemOpen
  \bibfield  {author} {\bibinfo {author} {\bibfnamefont {T.}~\bibnamefont
  {Vojta}}\ and\ \bibinfo {author} {\bibfnamefont {J.~A.}\ \bibnamefont
  {Hoyos}},\ }\href@noop {} {\bibfield  {journal} {\bibinfo  {journal} {Phys.
  Rev. Lett.}\ }\textbf {\bibinfo {volume} {112}},\ \bibinfo {pages} {075702}
  (\bibinfo {year} {2014})}\BibitemShut {NoStop}%
\bibitem [{\citenamefont {Spivak}\ \emph {et~al.}(2008)\citenamefont {Spivak},
  \citenamefont {Oreto},\ and\ \citenamefont {Kivelson}}]{Spivak}%
  \BibitemOpen
  \bibfield  {author} {\bibinfo {author} {\bibfnamefont {B.}~\bibnamefont
  {Spivak}}, \bibinfo {author} {\bibfnamefont {P.}~\bibnamefont {Oreto}}, \
  and\ \bibinfo {author} {\bibfnamefont {S.~A.}\ \bibnamefont {Kivelson}},\
  }\href {\doibase 10.1103/PhysRevB.77.214523} {\bibfield  {journal} {\bibinfo
  {journal} {Phys. Rev. B}\ }\textbf {\bibinfo {volume} {77}},\ \bibinfo
  {pages} {214523} (\bibinfo {year} {2008})}\BibitemShut {NoStop}%
\bibitem [{\citenamefont {Millis}\ \emph {et~al.}(2002)\citenamefont {Millis},
  \citenamefont {Morr},\ and\ \citenamefont {Schmalian}}]{Millis}%
  \BibitemOpen
  \bibfield  {author} {\bibinfo {author} {\bibfnamefont {A.~J.}\ \bibnamefont
  {Millis}}, \bibinfo {author} {\bibfnamefont {D.~K.}\ \bibnamefont {Morr}}, \
  and\ \bibinfo {author} {\bibfnamefont {J.}~\bibnamefont {Schmalian}},\ }\href
  {\doibase 10.1103/PhysRevB.66.174433} {\bibfield  {journal} {\bibinfo
  {journal} {Phys. Rev. B}\ }\textbf {\bibinfo {volume} {66}},\ \bibinfo
  {pages} {174433} (\bibinfo {year} {2002})}\BibitemShut {NoStop}%
\bibitem [{\citenamefont {Harris}(1974)}]{Harris}%
  \BibitemOpen
  \bibfield  {author} {\bibinfo {author} {\bibfnamefont {A.~B.}\ \bibnamefont
  {Harris}},\ }\href@noop {} {\bibfield  {journal} {\bibinfo  {journal} {J.
  Phys. C Solid State Physics}\ }\textbf {\bibinfo {volume} {7}},\ \bibinfo
  {pages} {1671} (\bibinfo {year} {1974})}\BibitemShut {NoStop}%
\bibitem [{\citenamefont {Sachdev}\ \emph {et~al.}(2004)\citenamefont
  {Sachdev}, \citenamefont {Werner},\ and\ \citenamefont {Troyer}}]{Sachd}%
  \BibitemOpen
  \bibfield  {author} {\bibinfo {author} {\bibfnamefont {S.}~\bibnamefont
  {Sachdev}}, \bibinfo {author} {\bibfnamefont {P.}~\bibnamefont {Werner}}, \
  and\ \bibinfo {author} {\bibfnamefont {M.}~\bibnamefont {Troyer}},\
  }\href@noop {} {\bibfield  {journal} {\bibinfo  {journal} {Phys. Rev. Lett.}\
  }\textbf {\bibinfo {volume} {92}},\ \bibinfo {pages} {237003} (\bibinfo
  {year} {2004})}\BibitemShut {NoStop}%
\end{thebibliography}%

\end{document}